\documentclass[useAMS,usenatbib]{mn2e}

\usepackage{txfonts,graphicx,amssymb,rotating}

\title[Disc photoevaporation I: hydrodynamic models]{Photoevaporation of protoplanetary discs I: hydrodynamic models}

\author[R.D.~Alexander, C.J.~Clarke \& J.E.~Pringle]
  {R.D.~Alexander$^{1,2,}$\thanks{email: rda@jilau1.colorado.edu},
  C.J.~Clarke$^1$
  and J.E.~Pringle$^1$ \\
   $^1$ Institute of Astronomy, Madingley Road, Cambridge, CB3 0HA, UK\\
   $^2$ JILA, 440 UCB, University of Colorado, Boulder, CO 80309-0440, USA}

\begin{document}

\pagerange{\pageref{firstpage}--\pageref{lastpage}} \pubyear{2006}

\maketitle

\label{firstpage}

\begin{abstract}
In this paper we consider the effect of the direct ionizing stellar radiation field on the evolution of protoplanetary discs subject to photoevaporative winds.  We suggest that models which combine viscous evolution with photoevaporation of the disc \citep*[e.g.][]{cc01} incorrectly neglect the direct field after the inner disc has drained, at late times in the evolution.  We construct models of the photoevaporative wind produced by the direct field, first using simple analytic arguments and later using detailed numerical hydrodynamics.  We find that the wind produced by the direct field at late times is much larger than has previously been assumed, and we show that the mass-loss rate scales as $R_{\mathrm {in}}^{1/2}$ (where $R_{\mathrm {in}}$ is the radius of the instantaneous inner disc edge).  We suggest that this result has important consequences for theories of disc evolution, and go on to consider the effects of this result on disc evolution in detail in a companion paper \citep*{paper2}.
\end{abstract}



\section{Introduction}
The evolution and eventual dispersal of discs around young stars is an important area of study, for theories of both star and planet formation.  It is now well-established that at an age of $\sim 10^6$yr most stars are surrounded by discs that are optically thick at optical and infrared wavelengths \citep[e.g.][]{kh95} and typically have masses of a few percent of a solar mass \citep[e.g.][]{beckwith90}.  However at an age of $\sim 10^7$yr most stars are not seen to have discs, suggesting that disc lifetimes are typically a few Myr \citep*[e.g.][]{haisch01}.  How stars lose their discs remains an unsolved question.  Observations of T Tauri stars (TTs) in the infrared and at millimetre wavelengths detect few objects intermediate between the disc-bearing classical T Tauri (CTT) and disc-less weak-lined T Tauri (WTT) states \citep*{skrutskie90,kh95,persi00,aw05}.  This suggests that the transition from CTT to WTT is very rapid, with the dispersal time estimated to be $\sim 10^5$yr \citep{sp95,ww96}. Moreover the simultaneous decline is disc emission across a wide range in wavelength suggests that the dispersal is essentially simultaneous across the entire radial extent of the disc. 

This ``two-time-scale'' behaviour is inconsistent with conventional models of disc evolution, which predict power-law declines in the disc and therefore predict dispersal times comparable to the disc lifetimes \citep*{hcga98,act99}.  However it has been shown that models which combine photoevaporation of the disc material with viscous evolution can reproduce this two-time-scale behaviour \citep*{cc01}.  Models of so-called photoevaporative disc winds were first suggested at least 20 years ago \citep[e.g.][]{bs82}.  However the first models which treated the flow and radiative transfer in a fully self-consistent manner were reported by \citet*{holl93} and \citet*{shu93}, and then extended in depth by \citet{holl94}.  Further studies have extended the study of the details in a number of ways (e.g.~the effects of dust, \citealt{ry97}; the inclusion of external radiation fields and non-ionizing UV radiation, \citealt*{jhb98}; detailed numerical hydrodynamics, \citealt{font04}), but the underlying principles remain the same.  Ionizing radiation from the central star produces a hot ionized layer on the surface of the disc, with conditions akin to an H\,{\sc ii} region.  Near to the star the ionized gas remains bound to the disc, but outside some critical radius (known as the gravitational radius and denoted by $R_{\mathrm g}$) the ionized layer becomes unbound and flows from the disc surface as a wind.  For solar-mass stars such as TTs the gravitational radius is typically a few AU.  For a constant ionizing flux $\Phi$ the wind rate is constant, as long as the disc remains optically thick to ionizing photons perpendicular to the plane of the disc.  Furthermore the diffuse (recombination) radiation field is the dominant source of ionizations at all radii of interest\footnote{\citet{holl94} found that the radius beyond which the diffuse field dominates is $3.8 \left(M_*/1\mathrm M_{\odot}\right)^{1/2} \left(R_*/1\mathrm R_{\odot}\right)^{2/3} \mathrm R_{\odot}$, which is much smaller than $R_{\mathrm g}$ for parameters typical of TTs.}.  In their ``weak-wind'' case, applicable to TTs, \citet{holl94} found that the mass-loss rate due to photoevaporation is given by
\begin{equation}\label{eq:wind_rate}
\dot{M}_{\mathrm {wind}} \simeq 4.4\times 10^{-10} \left(\frac{\Phi}{10^{41}\mathrm s^{-1}}\right)^{1/2} \left(\frac{M_*}{1\mathrm M_{\odot}}\right)^{1/2} \mathrm M_{\odot} \mathrm {yr}^{-1}.
\end{equation}
More recently, hydrodynamic modelling of a photoevaporative wind has resulted in slight modification of this result.  When hydrodynamic effect are considered the ``effective $R_{\mathrm g}$'' is reduced by a factor of 5 \citep{liffman03,font04}, and the mass-loss rate is reduced by a factor of around 3 \citep{font04}.  However the qualitative behaviour is unchanged from that of \citet{holl94}.

The so-called ``UV-switch'' model of \citet{cc01} couples a photoevaporative wind to a simple disc evolution model.  At early times the accretion rate through the disc is much larger than the wind rate, and the wind has a negligible effect.  However at late times photoevaporation becomes important, depriving the disc of resupply inside $R_{\mathrm g}$.  At this point the inner disc drains on its own, short, viscous time-scale, giving a dispersal time much shorter than the disc lifetime.  A number of similar studies have now been conducted \citep*{mjh03,acp03,ruden04,tcl05}, and this class of models shows a number of attractive properties.

However \citet{cc01} highlighted two key problems with the model.  Firstly, the model requires that TTs produce a rather large ionizing flux (of order $10^{41}$ ionizing photons per second), and secondly that the outer disc, beyond $R_{\mathrm g}$ is dispersed much too slowly to satisfy millimetre observations of TT discs.  We have previously shown that it is reasonable to treat TT chromospheres as having a constant ionizing flux in the range $\sim 10^{41}$--$10^{44}$s$^{-1}$ \citep*{chrom}.  We now seek to address the ``outer disc problem'' by highlighting a flaw in the original UV-switch model.  The photoevaporative wind models described above are based on the premise that the disc is optically thick to Lyman continuum photons at all radii.  Consequently the diffuse (recombination) radiation field is the dominant source of ionizing photons at large radii \citep{holl94}, and drives the photoevaporative wind.  However once the inner disc is drained it becomes optically thin to Lyman continuum photons, eliminating this diffuse field and rendering the simple wind prescription used above invalid.  Instead, once the inner disc has drained we must consider direct ionization of the inner edge of the disc.  \citet{cc01} neglect this process, and find that the time-scale for dispersal of the outer disc is limited by the time material takes to diffuse inward to $R_{\mathrm g}$ (as the most of the mass-loss occurs close to $R_{\mathrm g}$).  Further, the $R^{-5/2}$ dependence of the wind profile at large radii \citep{holl94} means that the mass-loss rate due to photoevaporation decreases significantly with time as the inner edge of the disc moves outward.  Consequently the dispersal of the outer disc occurs on the viscous time-scale of the {\it outer} disc, and thus the outer disc is dispersed in a time comparable to the disc lifetime, much too slowly to satisfy observational constraints.  We suggest that photoevaporation by the direct radiation field results in a dispersal time significantly shorter than that predicted by \citet{cc01}, and now seek to model the effects of this process on the evolution of the outer disc.

In this paper we model the photoevaporative wind produced by the direct field.  In Section \ref{sec:analytic_models} we present an analytic treatment of the problem, highlighting the important physical processes.  In Section \ref{sec:hydro_models} we extend our analysis by using numerical hydrodynamics, and describe the numerical simulations we have performed.  In Section \ref{sec:res} we present our results, and show that the wind profile is well-matched by a simple analytic form.  In Section \ref{sec:disc} we discuss the caveats that apply to our modelling, as well as the consequences of our results for disc evolution models, and in Section \ref{sec:summary} we summarize our conclusions.  We consider the effects of our results on disc evolution models in detail in a companion paper \citep[][hereafter Paper II]{paper2}.


\section{Analytic model}\label{sec:analytic_models}
\begin{figure}
        \begin{center}
        \resizebox{\hsize}{!}{
        \includegraphics{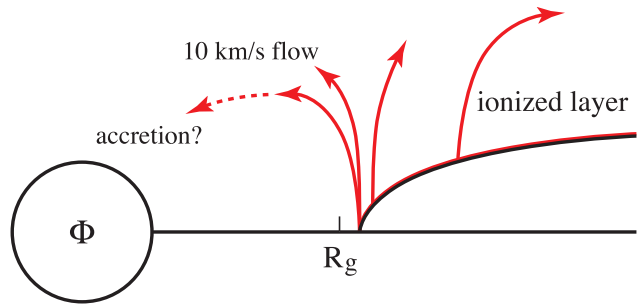}
        }
        \end{center}    
        \caption[Schematic picture of direct photoevaporation]{Schematic representation of the disc wind produced by direct photoevaporation.  Ionizing radiation from the star creates an ionized layer on the disc surface.  The ionized gas is unbound, and flows away from the disc at approximately the sound speed.}
        \label{fig:direct_schematic}
\end{figure}

We now consider the problem which we refer to as ``direct photoevaporation''.  As discussed above, after the inner-disc draining of the UV-switch model the inner disc becomes optically thin to Lyman continuum photons, and so we must consider the influence of the direct radiation field rather than the diffuse field considered previously.  We note that while the UV-switch model shows that the gas in the inner disc will become optically thin to ionizing radiation, some dust opacity may remain after the time at which the inner gas disc drains.  An detailed treatment of gas-grain dynamics is beyond the scope of this work, but we will consider this issue qualitatively in the discussion (Section \ref{sec:disc}).
We expect similar physics to apply in the case of the direct field as in the diffuse field case, with the ionizing radiation creating a thin ionized layer on the surface of the disc (see Fig.\ref{fig:direct_schematic}).  However the disc is truncated at some inner radius such that the ionized layer is unbound, and flows away from the surface as a disc wind.  Additionally, at the inner edge any flow perpendicular to the disc surface will move inward and so some material may be accreted on to the star (although it must presumably lose angular momentum to do so).  Obviously this is a complicated dynamic process, and in the next section we use numerical hydrodynamics to model the dynamics of direct photoevaporation.  However we first seek a theoretical framework in which to place these hydrodynamic models.  

In order to study the problem of direct photoevaporation analytically we use a similar approach to that adopted by \citet{holl94}.  We assume that the mass-loss per unit area from the disc at a given point is given by $\rho c_{\mathrm s}$, where $\rho$ is the density at the base of the region layer and $c_{\mathrm s}$ is the sound speed of the ionized gas.  Consequently the density at the at the base of the ionized region is critical to the determination of the photoevaporative wind.

\begin{figure}
        \begin{center}
        \resizebox{\hsize}{!}{
        \includegraphics{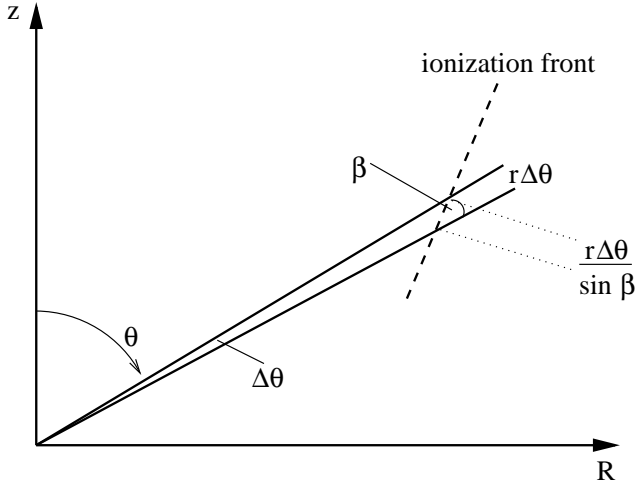}
        }
        \end{center}    
        \caption[Geometry of direct ionization problem]{Diagram showing geometry of the direct ionization problem.  Along a line-of-sight with polar angle $\theta$ and angular size $\Delta \theta$, the angle between the ionization front and the line-of-sight is $\beta$.  Consequently the projected length of the column perpendicular to the front is $r\Delta \theta/\sin\beta$.}
        \label{fig:geom_fig}
\end{figure}

We solve for the number density at the base of the ionized region (i.e.~at the ionization front) as follows.  We neglect recombinations between the source and the disc surface, and assume that the location of ionization front is determined by ionization balance only.  We also assume azimuthal symmetry, and integrate over the azimuthal coordinate throughout.  Consequently, along any given line-of-sight from the source the rate of recombinations, $N_{\mathrm {rec}}$, in a volume $\Delta V$ at the ionization front must balance the rate of ionizing photons absorbed at the front, $N_{\mathrm {ion}}$.  A column with polar angle $\theta$ and angular size $\Delta\theta$ has a total area equal to $2\pi r^2 \sin \theta \Delta \theta$, so for an ionizing flux $\Phi$ the ionization rate at the front is\footnote{At various points in this paper it is convenient to work in either cylindrical or spherical polar coordinates.  To avoid confusion we adopt the notation $(r,\theta,\phi)$ for spherical coordinates and $(R,z,\phi)$ for cylindrical.  Thus upper-case $R$ always represents cylindrical radius, while lower-case $r$ denotes spherical radius.}
\begin{equation}
N_{\mathrm {ion}} = \frac{1}{2}\sin \theta \Delta \theta \Phi \, .
\end{equation}
The geometry of the problem is shown in Fig.\ref{fig:geom_fig}.  We have assumed that at the ionization front we have ionization balance within some volume $\Delta V$.  Perturbations to the disc structure on a length-scale shorter than the disc scale-height $H(R)$ are unlikely to be dynamically stable \citep*[e.g.][]{lpf85}, so we assume that $\Delta V$ has a thickness (perpendicular to the front) that is equal to the disc scale-height $H(R)$.  We verify this assumption {\it a posteriori} in Section \ref{sec:inner_edge}.  Consequently $\Delta V$ is given by
\begin{equation}
\Delta V = 2\pi R H \frac{r \Delta \theta}{\sin \beta} \, ,
\end{equation}
where $\beta$ is the angle between the ray-path and the ionization front (see Fig.\ref{fig:geom_fig}). 
The spherical radius $r$ is related to the cylindrical radius $R$ by $R=r\sin\theta$, and so the recombination rate is
\begin{equation}
N_{\mathrm {rec}} = 2\pi \alpha n_0^2 R H \frac{R \Delta \theta}{\sin \theta \sin \beta} \, ,
\end{equation}
where $n_0(R)$ is the number density at the ionization front and $\alpha$ is the hydrogen recombination coefficient.  If we equate the ionization and recombination rates and re-arrange we find that
\begin{equation}\label{eq:base_den}
n_0(R) = \left(\frac{\Phi \sin^2 \theta \sin \beta}{4\pi \alpha R^2 H(R)}\right)^{1/2} \, .
\end{equation}
In general, therefore, the density at the ionization front at a given radius $R$ depends on the geometry of the ionization front.  This in turn depends of the density structure of the unperturbed disc, and the form of the density profile does not permit exact analytic solutions.  However we can look at the behaviour of the density in specific cases.  For example, along the midplane ($\theta = \pi/2$) we expect the ionization front to be perpendicular to the line-of-sight to the source, and so $\sin \beta = 1$.  In most discs $H(R)$ is an increasing function of $R$, and at radii beyond the inner disc edge we have 
$\sin \theta < 1$ and $\sin \beta < 1$, so $n_0(R)$ falls off more steeply than $R^{-3/2}$
The exact profile of the base density is determined by the geometry of the ionization front, which must be evaluated numerically.  However before we consider numerical modelling of the wind it is instructive to frame the problem in terms of known parameters and unknown scaling constants.

We note at this point that the form we have derived for $n_0(R)$ (Equation \ref{eq:base_den}) is qualitatively similar to the ``strong wind'' solution derived by \citet{holl94}.  In this case a strong stellar wind modifies the disc structure out to a radius beyond $R_{\mathrm g}$, "driving down" the disc atmosphere and enabling radiation to penetrate to lower heights above the disc midplane at larger radii.  Consequently the $n_0 \propto R^{-3/2}$ scaling, which applies only to the inner disc in the weak-wind case, extends well beyond $R_{\mathrm g}$ in the presence of a strong stellar wind.  Such a wind is not found for TTs, and instead this case is more readily applicable to O- and B-type stars.   However the geometry of the radiative transfer problem is similar to that considered here, with a similar lack of a disc atmosphere at small radii.  The numerical scaling will differ, as even in the strong wind case the diffuse field still dominates, but we expect the form of the mass loss due to the direct field to be qualitatively similar to that derived by \citet{holl94} for the case of a strong stellar wind.

\subsection{Scaling parameters}
Following \citet{holl94}, we evaluate the wind mass-loss rate per unit area as
\begin{equation}
\dot{\Sigma}_{\mathrm {wind}}(R,t) = 2 \mu m_{\mathrm H}n_0(R,t) u_{\mathrm {l}}(R,t)  \, ,
\end{equation}
where $\mu$ is the mean molecular weight of the gas, $m_{\mathrm H}$ is the mass of a hydrogen atom, and the factor of 2 accounts for mass-loss from both sides of the disc.  The launch velocity $u_{\mathrm {l}}(R,t)$ is of order the sound speed of the ionized gas.
For the moment we make the simplifying assumption that at the inner disc edge $\Delta V$ has a radial thickness comparable to the vertical scale-height.  As noted above, along the midplane we have $\sin \theta = \sin \beta = 1$, and so the inner edge density $n_{\mathrm {in}}$ can be expressed as
\begin{equation}\label{eq:n_in}
n_{\mathrm {in}} = C \left(\frac{\Phi}{4\pi \alpha (H/R)_{\mathrm {in}} R_{\mathrm {in}}^3}\right)^{1/2} \, ,
\end{equation}
where $R_{\mathrm {in}}$ is the radius of the inner disc edge and $C$ is an order-of-unity scaling constant.  We further assume that the base density $n_0(R)$ (Equation \ref{eq:base_den}) can be related to the inner edge density $n_{\mathrm {in}}$ by a dimensionless shape function $f\left(R/R_{\mathrm {in}}\right)$, which depends on the disc structure (and therefore depends on the form of $H(R)$).  Therefore
\begin{equation}
n_0(R) = n_{\mathrm {in}} f\left(\frac{R}{R_{\mathrm {in}}}\right)
\end{equation}
and the launch velocity can be written as
\begin{equation}\label{eq:launch_vel}
u_{\mathrm {l}}(R) = D c_{\mathrm s} \, .
\end{equation}
Here $D$ is another order-of-unity scaling constant and $c_{\mathrm s}=10$km s$^{-1}$ is the sound speed of the ionized gas.  Therefore the mass-loss profile takes the form
\begin{equation}\label{eq:prof_form}
\dot{\Sigma}_{\mathrm {wind}}(R,t) = 2 CD \mu m_{\mathrm H} c_{\mathrm s}  n_{\mathrm {in}}(t) f\left(\frac{R}{R_{\mathrm {in}}(t)}\right)\, .
\end{equation}
The total mass-loss rate is found by integrating this from $R_{\mathrm {in}}$ to some outer radius $R_{\mathrm {out}}$:
\begin{equation}
\dot{M}(<R_{\mathrm {out}}) = \int_{R_{\mathrm {in}}}^{R_{\mathrm {out}}} 2\pi R \dot{\Sigma}_{\mathrm {wind}}(R) dR \, .
\end{equation}
Substituting for $n_{\mathrm {in}}$ from Equation \ref{eq:n_in} and integrating gives
\begin{eqnarray}
\dot{M}(<R_{\mathrm {out}}) = 4\pi CD \mu m_{\mathrm H} c_{\mathrm s} \left(\frac{\Phi}{4\pi \alpha (H/R)_{\mathrm {in}}}\right)^{1/2} R_{\mathrm {in}}^{1/2}
\nonumber \\
\times  \int_1^{R_{\mathrm {out}}/R_{\mathrm {in}}} x f(x) dx \, .
\end{eqnarray}
Typically, for TT discs, $R_{\mathrm {out}} \gg R_{\mathrm {in}}$.  Further, if we assume that the integral converges as $R_{\mathrm {out}}/ R_{\mathrm {in}} \rightarrow \infty$ (i.e.~that the form of $f(x)$ falls off faster than $x^{-2}$), then we see that the mass-loss rate from the wind scales as $R_{\mathrm {in}}^{1/2}$.  Consequently we expect the mass-loss rate to increase as the inner edge of the disc evolves outwards.  If we re-scale in terms of parameters typical of TTs (taking $(H/R)_{\mathrm {in}}=0.05$), we find 
\begin{eqnarray}\label{eq:mout_anal}
\dot{M}(<R_{\mathrm {out}}) = 1.74\times10^{-9} \, CD \, \mu \, \left(\frac{\Phi}{10^{41}\mathrm s^{-1}}\right)^{1/2}  \left(\frac{R_{\mathrm {in}}}{3\mathrm{AU}}\right)^{1/2}
\nonumber \\
\times  \int_1^{R_{\mathrm {out}}/R_{\mathrm {in}}} x f(x) dx \, \mathrm M_{\odot}\mathrm{yr}^{-1} \, .
\end{eqnarray}
As noted earlier, the shape function $f(x)$ depends on both the radial density profile and the form of $H(R)$.  We expect a profile that falls off faster than $R^{-3/2}$, but require numerical solution in order to determine the exact form of $f(x)$. Moreover we must also determine the scaling constants $C$ and $D$ numerically, so we turn to numerical simulations in order to pursue this problem further.


\section{Hydrodynamic model}\label{sec:hydro_models}
In this section we construct and evaluate hydrodynamic models of the disc wind produced by direct photoevaporation.  We first describe the computational method used to investigate the problem and the simulations used to study the behaviour of the wind.  We then present our results, which are framed in terms of the scaling parameters described above, before discussing the various caveats which apply to the models.

\subsection{Computational method}
To investigate the dynamics of direct photoevaporation we use the two-dimensional (2-D) grid-based hydrodynamics code {\sc zeus2d} \citep{sn92}.  {\sc zeus2d} uses operator splitting to integrate the hydrodynamic equations forward in time.  Each equation is split into parts, which are then evaluated consecutively using the results from the previous evaluation.  Further splitting is done between the two grid directions, with the order alternating on consecutive timesteps.  The hydrodynamic equations are split into source and transport terms.  The source terms are evaluated first, and the new values are then transported by computing fluxes across cell faces.  The form of this computation makes it relatively straightforward to add heating due to ionizing radiation, as it is easy to update the local energy at the end of each timestep, before subsequent evaluation of the source terms.  

{\sc zeus2d} provides several alternative options for the interpolation scheme and artificial viscosity.  We adopt the standard van Leer (second-order) interpolation scheme, and the standard von Neumann \& Richtmyer form for the artificial viscosity (with $q_{\mathrm {visc}}=2.0$).  In evaluating the timestep we adopt a Courant number of 0.4.  Additionally, we adopt the ideal gas equation of state 
\begin{equation}
p = (\gamma - 1) e \, 
\end{equation}
throughout, where $p$ is gas pressure, $e$ is energy density (per unit volume)\footnote{Note that for computational reasons {\sc zeus2d} works with the energy density per unit volume, denoted by $e$, rather than the more conventional energy per unit mass.  Here $e$ is equal to the quantity denoted by ``$\rho e$'' in most hydrodynamic texts.}, and $\gamma=5/3$ is the adiabatic exponent.

\subsubsection{Ionizing radiation}
In order to model the effects of ionizing radiation in {\sc zeus2d} it is necessary to simplify the radiative transfer problem, as it is not practical to make a full radiative transfer calculation at each timestep.  In order to solve the problem we assume that the only significant heat source is the absorption of Lyman continuum photons, and that the only significant coolant is the re-emission of such photons by radiative recombination.  Consequently the problem of evaluating the heating and cooling due to radiation is reduced to a problem of computing ionization balance.  We assume that the gas is either extremely optically thick to ionizing photons produced by recombination, in which case they are absorbed locally (Case B recombination), or assume that the gas is extremely optically thin to recombination photons, in which case they escape from the system (Case A).  Consequently recombination photons can be neglected in the computation of ionization balance.  This approximation (specifically the assumption of local absorption) is known as the ``on-the-spot'' (henceforth OTS) approximation.  Where the OTS approximation is not valid is the case where ionizing photons produced by recombinations travel some distance before being re-absorbed.  We discuss the validity of the OTS approximation {\it a posteriori} in Section \ref{sec:disc}.

In order to model the effects of a central ionizing source we have added two routines to the code.  These routines require that a polar [i.e.~$(r,\theta)$] coordinate grid is used, so that the radial columns represent ray paths from the central source.  The first routine is performed at the end of each timestep (i.e.~after the transport step), and solves for ionization balance along each ray path in order to find the location of the ionization front.  It then sets a flag array to indicate whether cells should be ionized, neutral or ``boundary''.  To do this we solve the equation of ionization balance along each radial grid column (denoted by subscript $_j$)
\begin{equation}\label{eq:ioniz_rate}
\frac{dN_j}{dt} = \frac{1}{2}\sin \theta_j \Delta \theta_j \, \Phi - \sum_j \alpha f_{i,j} n_{i,j}^2 V_{i,j} \, ,
\end{equation}
where $N_j$ is the total number of ionized atoms in the along the $j$th radial grid column, $\alpha$ the recombination rate coefficient, $n_{i,j}$ the particle number density at cell $(i,j)$, and $V_{i,j}$ ($= 2\pi \sin \theta_j \Delta r_i \Delta \theta_j$) the cell volume.  The term $f_{i,j}$ is the fraction of cell $(i,j)$ which is ionized, and is set to unity for ionized cells or zero for neutral cells.  In ``boundary'' cells it has an intermediate value (see Equation \ref{eq:f} below).  The particle number density is obtained from the mass density by assuming that the gas is entirely neutral hydrogen: $n_{i,j} = \rho_{i,j}/m_H$ (i.e.~$\mu=1$).  Essentially the first term on the right-hand side is the number of ionizations per unit time in grid column $j$, and the second term is the number of recombinations per unit time.  Thus the change in the number of ionized atoms in a single timestep of length $\Delta t$ is
\begin{equation}
\Delta N_j = \Delta t \left(\frac{1}{2}\sin \theta_j \Delta \theta_j \, \Phi - \sum_j \alpha f_{i,j} n_{i,j}^2 V_{i,j}\right) \, .
\end{equation}
Having evaluated this number we evaluate the updated number of ionized atoms along the column to be
\begin{equation}
N_j^{n+1} = N_j^n + \Delta N_j \, .
\end{equation}
We then evaluate the number of atoms enclosed at a given radius $r_i$ along the $j$th column as
\begin{equation}\label{eq:mass_enc}
N_{\mathrm {enc},i} =  \sum_{r_0}^{r_i} n_{i,j} V_{i,j} \, ,
\end{equation}
where $r_0$ is the inner grid radius.  If $N_j^{n+1} > N_{\mathrm {enc},i}$ then the cell is flagged as ``ionized'', and if $N_j^{n+1} < N_{\mathrm {enc},i}$ then the cell is flagged as ``neutral''.  The cell along each ray path where equality is reached is flagged as the ``boundary cell''.  In this cell we evaluate the fraction, $f_{i,j}$, of the cell which must be ionized so that $N_j^{n+1} = N_{\mathrm {enc}}$:
\begin{equation}\label{eq:f}
f_{i,j} = \frac{N_j^{n+1} - N_{\mathrm {enc},i-1}}{n_{i,j} V_{i,j}} \, .
\end{equation}
In addition, a check is performed to ensure that the recombination time-scale at each grid cell which is ``ionized'', $t_{\mathrm {rec}} = (\alpha n)^{-1}$, is much longer than the dynamic timestep.  As long as this condition is not violated we can safely neglect radiative cooling of the ionized gas.  The density of the ionized material is sufficiently small that this is not a concern in any of our simulations.

Following this we have added another subroutine, which adjusts the energy of the gas according to the values the ionization flags calculated as above.  ``Ionized'' gas is treated as isothermal with a sound speed of 10kms$^{-1}$, ``neutral'' gas is left unchanged.  Thus at each cell the energy is adjusted as follows:
\begin{equation}\label{eq:EoS}
e_{i,j}^{n+1} = \left\{ \begin{array}{ll}
  e_{i,j}^{n} & \textrm{if ``neutral''}\\
  (c_{\mathrm {bc}}^2 \rho_{i,j})/(\gamma -1) & \textrm{if ``boundary''}\\ 
  (c_{\mathrm {hot}}^2 \rho_{i,j})/(\gamma -1) & \textrm{if ``ionized''}\\ 
\end{array} \right.
\end{equation}
where $c_{\mathrm {hot}}=10$kms$^{-1}$, and the sound speed in the boundary cell $c_{\mathrm {bc}}$ is evaluated as
\begin{equation}\label{eq:c_bc}
c_{\mathrm {bc}}^2 = f_{i,j} c_{\mathrm {hot}}^2 + (1-f_{i,j})c_{\mathrm {cold}}^2 \, .
\end{equation}
In this manner the energy in the boundary cells is evaluated is as the weighted mean of the ``cold'' and ``hot'' energies.  $c_{\mathrm {cold}}$ is evaluated as $c_{\mathrm {cold}}^2=P/\rho$ on the timestep when a cell first becomes partially ``ionized''.  The values of $f_{i,j}$ and $c_{\mathrm {bc}}$ are carried to the following timestep so that this value can be recovered from Equation \ref{eq:c_bc} if the ionization front remains in the same cell for one or more timesteps.

Lastly, in order to increase the performance of the code we have parallelised it, for a shared-memory architecture, using the OpenMP formalism\footnote{See {\tt http://www.openmp.org} .}.  Tests showed that the numerical results were identical to those obtained from running the code on a single processor, and that the parallelisation is around 85--90\% efficient when running on 4 processors.

\subsection{Testing the code}\label{sec:Spitzer_test}
To test the accuracy of this algorithm we consider the well-studied case of an H\,{\sc ii} region expanding into a uniform gas cloud of number density $n_0$.  This problem has a well-known 1-D analytic solution, first studied by \citet{spitzer78}.  If an ionization front expands spherically into a uniform gas cloud of density $n_0$, then the radius of the ionization front $r_{\mathrm i}(t)$ at time $t$ is
\begin{equation}\label{eq:Spitzer_soln}
r_{\mathrm i}(t) = r_{\mathrm s} \left(1+ \frac{7}{4}\frac{c_{\mathrm {hot}}t}{r_{\mathrm s}}\right)^{4/7}
\end{equation}
where $r_{\mathrm s}$ is the Str\"omgren radius
\begin{equation}
r_{\mathrm s} = \left(\frac{3 \Phi}{4\pi \alpha_B n_0^2}\right)^{1/3} \, .
\end{equation}
\begin{figure}
        \begin{center}
        \resizebox{\hsize}{!}{
        \includegraphics[angle=270]{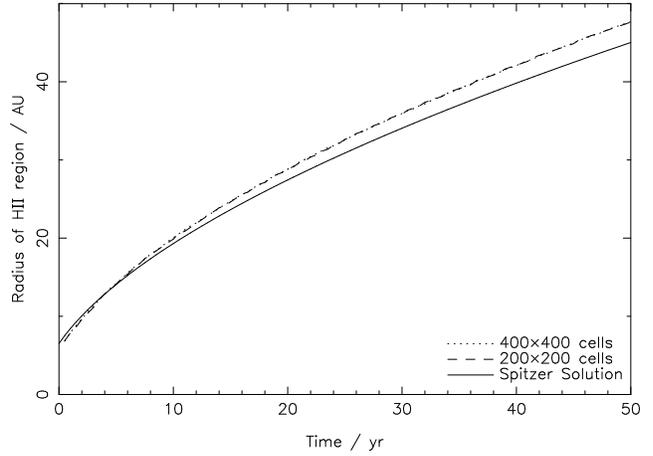}
        }
        \end{center}    
        \caption{Results of the test simulation, for the expansion of an H\,{\sc ii} region into a uniform density cloud.  Here the number density $n_0=10^6$cm$^{-3}$ and the ionizing flux $\Phi=10^{42}$s$^{-1}$.  The analytic (Spitzer) solution is plotted as a solid black line, with the numerical solutions using two different grid resolutions shown as dashed and dotted lines.}
        \label{fig:spittest}
\end{figure}
Here $\alpha_B$ is the Case B recombination coefficient for atomic hydrogen at $10^4$K, which has a value of $\alpha_B=2.6\times10^{-13}$cm$^3$s$^{-1}$ \citep{allen}.  We use a $200\times200$ cell grid covering the range $r=[0\mathrm{AU},50\mathrm{AU}]$ and $\theta = [0,\pi/2]$.  Our modified version of {\sc zeus2d} solves this problem numerically, and a comparison between the analytic and numerical solutions is shown in Fig.\ref{fig:spittest}.  The code reproduces both the initial Str\"omgren radius and the power-law rise well, accurate at the $\simeq 5$\% level.  Runs at higher resolution do not change the results significantly, indicating that the procedure is numerically converged.  We note however that this is not a particularly stringent test of the code for the conditions we wish to model, as ionization of a disc means that the density profile rises exponentially along ray paths.  Consequently we expect the ionization front to remain in a single grid cell for many more timesteps in a disc simulation than in this test, and the treatment of the boundary cell is the primary uncertainty in the computational algorithm. However in a spherical geometry the expansion of an H\,{\sc ii} region into a cloud with such a steeply rising density profile would naturally be Rayleigh-Taylor unstable, so this problem cannot be used to test the code.  In short, we find that our code performs well when considering the expansion of an H\,{\sc ii} region into a uniform density cloud, but note also that this is not an especially good test for the problem at hand.

\subsection{Disc model and initial conditions}
In order to model direct photoevaporation it is necessary to set up a stable disc configuration in {\sc zeus2d}.  We neglect magnetic fields, disc self-gravity and radiation hydrodynamics (these options are turned off in the code) and adopt an $(r,\theta)$ grid covering a polar angle of $\theta = [0,\pi/2]$ (i.e.~assuming symmetry and simulating only one quadrant of the disc).  The rotation option in the code, which introduces a centrifugal  ``pseudo-force'' is turned on, and accelerations due to gravity are evaluated using only a point mass placed at the origin.  The ionization subroutines are included, but initially the ionizing flux is set to zero so that they have no effect.  Both the upper and lower angular boundaries are set to be reflective in order to account for the symmetry of the problem, and both the inner and outer radial boundaries are set as ``outflow'' boundaries.  We address the influence of these boundary conditions later (see Section \ref{sec:numerics}).

To determine initial conditions we use a 1-D reference model of the form described by \citet[][ see also Paper II]{cc01}.  We use the updated wind profile of \citet[][ kindly provided in numerical form by Ian McCarthy]{font04} rather than the original profile of \citet{holl94} used by \citet{cc01}.  We expect direct photoevaporation to become significant once the inner disc is drained, so we run the 1-D evolution model to the point where the inner disc (inside the draining radius) becomes optically think to ionizing photons.  The reference model has an ionizing flux of $\Phi=10^{41}$s$^{-1}$, an initial disc mass of $0.05$M$_{\odot}$, and an initial accretion rate $5.0\times 10^{-7}$M$_{\odot}$yr$^{-1}$.  At the point when the inner disc is drained (at an age of 11.1Myr) the disc has a total mass of 0.001M$_{\odot}$, and in order to remove numerical fluctuations we use a functional fit to the density profile at this point.  The functional fit to the midplane density takes the form 
\begin{equation}\label{eq:func_form}
\rho(R,z=0) \propto (R - R_{\mathrm {in}})^{1/n} \frac{1}{R^{2+1/n}} \, .
\end{equation}
This is a $R^{-2}$ decline at large radii, with a cutoff at radii close to some inner edge radius $R_{\mathrm {in}}$.  $\rho(R,z=0) \propto R^{-2}$ is consistent with a $\Sigma \propto R^{-1}$ surface density and constant $H/R$.  This profile peaks at a radius $R_0$, with the parameters related by
\begin{equation}\label{eq:n_index}
n = \frac{R_{\mathrm {in}}}{2(R_0 - R_{\mathrm {in}})} \, .
\end{equation}
Thus $R_{\mathrm {in}}$ is the inner truncation radius and $R_0$ the radius at which the midplane density peaks.  Fitting this profile to the reference model gives $R_{\mathrm {in}} = 2.25$AU and $R_0 = 8.25$AU, and therefore $n=0.1875$.

Consequently the input parameters to the disc model are as follows: $M_*$ (stellar mass), $\Sigma_0$ (surface density at $R=R_0$), $R_{\mathrm {in}}$, $R_0$, $H/R$ and ionizing flux $\Phi$.  The initial disc structure is obtained from these parameters as follows.  The reference midplane density $\rho(R=R_0,z=0)$ is evaluated as
\begin{equation}
\rho(R=R_0,z=0) = \frac{\Sigma_0}{\sqrt{2\pi} H_0} \, ,
\end{equation}
where the scale height at $R_0$ is $H_0=R_0 \, (H/R)$.  This fixes the constant of proportionality demanded in Equation \ref{eq:func_form}, and the cutoff index $n$ is evaluated as specified in Equation \ref{eq:n_index}.  Consequently, at every point on the $(r,\theta)$ grid the mass density and energy density are set thus\footnote{Note that the polar and cylindrical coordinates are related by $R=r \sin \theta$ and $z=r\cos \theta$.}:
\begin{equation}
\rho(r,\theta) = \rho(R,z=0) \exp\left(-\frac{z^2}{2H^2}\right)
\end{equation}
and 
\begin{equation}
e(r,\theta) = \frac{1}{\gamma-1} \rho(R,z=0) \frac{GM_*}{R} \left(\frac{H}{R}\right)^2 \exp\left(-\frac{z^2}{2H^2}\right) \, .
\end{equation}
The functional form is undefined for $R \le R_{\mathrm {in}}$, and so the density and energy are set to a constant, small value in this region.  (We use a value that is $10^{-15}$ times the maximum value.)  The radial and polar velocities are set to zero.  The rotational velocity is set to the Keplerian value, with small corrections made to balance the radial pressure gradient arising from the choice of density profile.  

Throughout our simulations we adopt a grid which is linearly-spaced in both $r$ and $\theta$.  There are computational advantages to using logarithmic spacing in $r$ \citep[see e.g.][]{bate02}, in particular the fact that the grid cells can all be approximately square (i.e.~$\Delta r \simeq r\Delta \theta$).  However a logarithmic grid naturally concentrates the highest resolution close to the inner boundary, while we seek to place the inner disc edge at larger radius.  Also, our simulations do not cover a large dynamic range in radius (typically only around one order of magnitude).  Consequently we find that a logarithmic grid requires significantly more grid cells than a linear one in order to achieve the same resolution at the inner disc edge, resulting in significantly larger CPU requirements, so we adopt a linear grid throughout.

\subsection{Simulations}
\begin{table*}
 \centering
  \begin{tabular}{ccccccccc}
  \hline
Simulation & $N_r$ & $N_{\theta}$ & [$r_{\mathrm{min}}$,$r_{\mathrm{max}}$] & $R_{\mathrm {in}}$ & $R_0$ & $\Sigma_0$ & $H/R$ & $\Phi$ \\
  &  &  &  AU  & AU & AU & 10$^{-2}$g cm$^{-2}$ &  & $10^{41}$s$^{-1}$ \\

  \hline
{\sc Reference} & 400 & 200 & [1.0,9.0] & 2.25 & 8.25 & 4.46 & 0.05 & 0.0 \\
{\sc Edge1} & 400 & 200 & [1.0,9.0] & 2.25 & 8.25 & 4.46 & 0.05 & 1.0 \\
{\sc Edge2} & 400 & 200 & [1.0,9.0] & 2.25 & 8.25 & 4.46 & 0.1 & 1.0 \\
{\sc Edge3} & 400 & 200 & [1.0,9.0] & 2.25 & 8.25 & 4.46 & 0.05 & 10.0 \\
{\sc ConvTest} & 800 & 400 & [1.0,9.0] & 2.25 & 8.25 & 4.46 & 0.05 & 1.0 \\
{\sc Boundary} & 800 & 200 & [1.0,17.0] & 2.25 & 8.25 & 4.46 & 0.05 & 1.0 \\
{\sc Profile1} & 1200 & 100 & [1.0,49.0] & 2.25 & 8.25 & 4.46 & 0.05 & 1.0 \\
{\sc Profile2} & 1200 & 100 & [1.0,49.0] & 2.25 & 8.25 & 4.46 & 0.1 & 1.0 \\
{\sc Profile3} & 1200 & 100 & [1.0,49.0] & 2.25 & 8.25 & 4.46 & 0.05 & 10.0 \\
{\sc Profile4} & 1200 & 100 & [1.0,49.0] & 2.25 & 8.25 & 4.46 & 0.075 & 1.0 \\
{\sc LaunchVel} & 1200 & 100 & [1.0,49.0] & 13.5 & 20.0 & 10.80 & 0.05 & 1.0 \\

\hline
\end{tabular}
\vspace*{16pt}
  \caption[List of simulations]{List of simulations run, showing resolution and physical parameters for each simulation.  $N_r$ and $N_{\theta}$ are the number of grid cells used in the radial and angular dimensions respectively.}\label{tab:sims}
\end{table*}
As seen above, all of our hydrodynamic disc models are specified by the five input parameters $M_*$, $\Sigma_0$, $R_{\mathrm {in}}$, $R_0$ and $H/R$, plus the ionizing flux $\Phi$.  We adopt $M_*=1\mathrm M_{\odot}$ throughout.  A number of simulations were run, using the parameters specified in Table \ref{tab:sims}.  The majority of these simulations use values of $\Sigma_0$, $R_{\mathrm {in}}$ and $R_0$ from the reference model.  However these parameters were also varied in specific cases.  Our fiducial model adopts $H/R=0.05$ and $\Phi=10^{41}$s$^{-1}$, and we use a suite of simulations to study the effects of varying these parameters.  
All of the simulations were run for many outer orbital times beyond the point where the initial transients died out.
Here we summarize the simulations conducted:
\begin{description}
\item[{\sc Reference}] This simulation was run as a reference model to demonstrate stability of the initial conditions, with the ionizing flux $\Phi=0$.  The values of $\Sigma_0$, $R_{\mathrm {in}}$ and $R_0$ were taken from the reference model.  We adopt the fiducial value of $H/R=0.05$.  The grid resolution was chosen so that $\Delta r=0.02$AU and $\Delta \theta = \frac{\pi}{400}$.  With these choices the grid cells are approximately square (i.e.~$\Delta r \simeq r\Delta \theta$) at $r=2.5$AU.  This ensured that the scale-height $H$ was resolved into $>5$ cells throughout.  In order to prevent the run-time becoming unreasonably long it was necessary to limit the radial range to [1.0AU,9.0AU].
\item[{\sc Edge}]  These simulations were run to study the evolution of the inner disc edge.  The fiducial model uses the same parameters as the reference model above, but with $\Phi=10^{41}$s$^{-1}$.  Further models were run with $H/R=0.1$ and $\Phi=10^{42}$s$^{-1}$ to investigate the effect of varying these parameters.  
\item[{\sc ConvTest}]  This model was run at double the resolution of the fiducial model in order to check that the code was numerically converged.
\item[{\sc Boundary}]  This model was run at the same resolution as the fiducial model but with double the radial range, in order to study the influence of the outer boundary on the results.
\item[{\sc Profile}] These simulations were run at slightly lower spatial resolution over a much larger radial range ([1.0AU,49AU]), in order to study the mass-loss profile far from the inner disc edge.  As with the inner edge simulations, 3 simulations were run: the fiducial model, $H/R=0.1$, and $\Phi=10^{42}$s$^{-1}$.  In practice it was found that the mass-loss profile was rather sensitive to the disc thickness, and so a further simulation was run with an intermediate value of $H/R=0.075$. 
\item[{\sc LaunchVel}] For this simulation the disc was ``moved'' to a larger radius ($R_0=20.0$AU), in order to study the launch velocity at the ionization front in the regime where the sound speed of the ionized gas was greater than the Keplerian velocity.

\end{description}

\section{Results}\label{sec:res}
The {\sc reference} model, where the ionizing flux was set to zero, demonstrates that the initial conditions are stable.  After several outer orbital times the only changes to the structure are a slight vertical expansion of the upper region (due to the failure of the $z \ll R$ approximation), and a slight spreading at the inner cutoff radius (due to a small pressure mismatch at the cutoff radius: the functional form has a discontinuity in $dP/dR$ here).  In addition there is some outflow (and also some spurious reflection) of material at the outer radial boundary, due to the artificial pressure gradient introduced at the boundary.  However we are satisfied that the initial configuration is stable, and now seek to study the effects of ionizing radiation on the disc.

At this point it is important to note a peculiarity inherent to grid-based codes such as {\sc zeus2d}.  When using a grid to specify hydrodynamic variables there are two distinct methods we can use to extract data from the simulations.  We can either consider the local values of the various hydrodynamic variables by looking at their values in individual grid cells, or we can study integrated properties across many cells.  We wish to study the behaviour of the disc at the ionization front, as the flow is determined by the hydrodynamic properties at the front.  As seen in Section \ref{sec:analytic_models}, this can be done either by looking at the density and launch velocity at the ionization front, or by considering the integrated mass-loss rates inside fixed radii.  In terms of the simulations, the density and velocity at the front are ``single-cell'' quantities, whilst the mass outflow rate is a ``many-cell'' quantity.  Both are useful, but in general in our simulations we find that the ``many-cell'' outflow rate is more robust against numerical fluctuations.  We also note that our treatment of the ionization front inherently results in the measurement of ``single-cell'' quantities near the ionization front being resolution-dependent, even if the simulation is converged.  This is because in order to measure properties at the front it is necessary to look at the first ``ionized'' cell: with a different cell size this is at a different distance from the front, and therefore always depends on the resolution.

\subsection{Flow solution}
\begin{figure}
\centering
        \resizebox{\hsize}{!}{
        \includegraphics[angle=90]{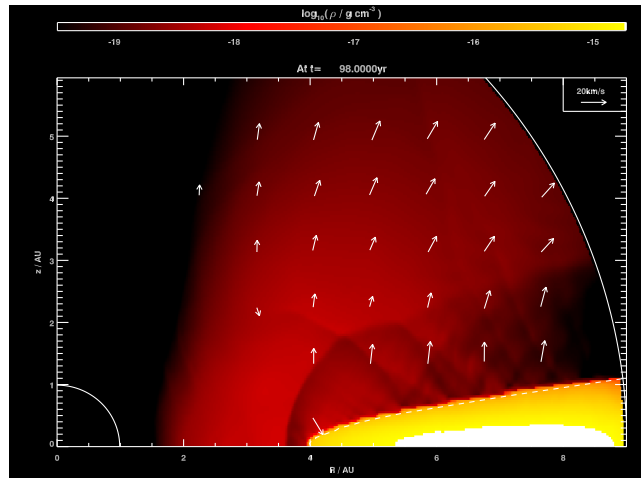}
        }
        \caption[Typical snapshot]{Snapshot of simulation {\sc Edge1} at $t=98$yr.  Density is plotted as a colour scale, with the grid boundaries denoted by solid lines and the ionization front by a dashed line.  Velocity vectors are plotted at regular intervals, but are omitted when they are either smaller than one-fifth the length of the reference vector, or when the density is below the minimum of the colour-scale.  Note, however, that the vectors represent the velocities at individual grid cells, and so can be prone to fluctuations.  Note also that the data is interpolated from the polar computational grid on to a rectangular coordinate grid in order to create the plot, so the resolution of the image does not reflect the resolution of the simulation.}
        \label{fig:snapshot}
\end{figure}

All of our models show a similar flow field. A snapshot of the fiducial {\sc Edge1} model is shown in Fig.\ref{fig:snapshot}.  The ionization front is perpendicular to the midplane at some inner edge radius, follows the disc surface as radius increases and becomes asymptotic to the line-of-sight from the ionizing source at large radii.  The inner edge radius advances slowly with time, with the consequence that the flow, while stable, is never entirely steady.  The flow is launched almost perpendicular to the ionization front.  Consequently at large radii the flow is launched near to vertically, while at the inner edge the launch velocity is radially inwards.  Conservation of angular momentum prevents the accretion of material through the inner radial boundary.   The ionized gas at the inner edge follows a ballistic trajectory, with the closest point of approach visible in the snapshots as a sharp decrease in the density of the ionized gas at approximately half the inner edge radius (as we would expect for material launched inward at close to the Keplerian speed).  Additionally, there is a shock front produced by the interaction between the outflowing and inflowing gas near to the inner disc edge, which again is clearly visible in the snapshots as a discontinuity in the density of the ionized gas.  However the flow between the disc surface and the shock front is supersonic, so the shock front does not influence the flow at the ionization front.  As the flow progresses along streamlines it becomes almost radial, but with streamlines which diverge from the inner disc edge rather than the origin.

\begin{figure}
\centering
        \resizebox{\hsize}{!}{
        \includegraphics[angle=90]{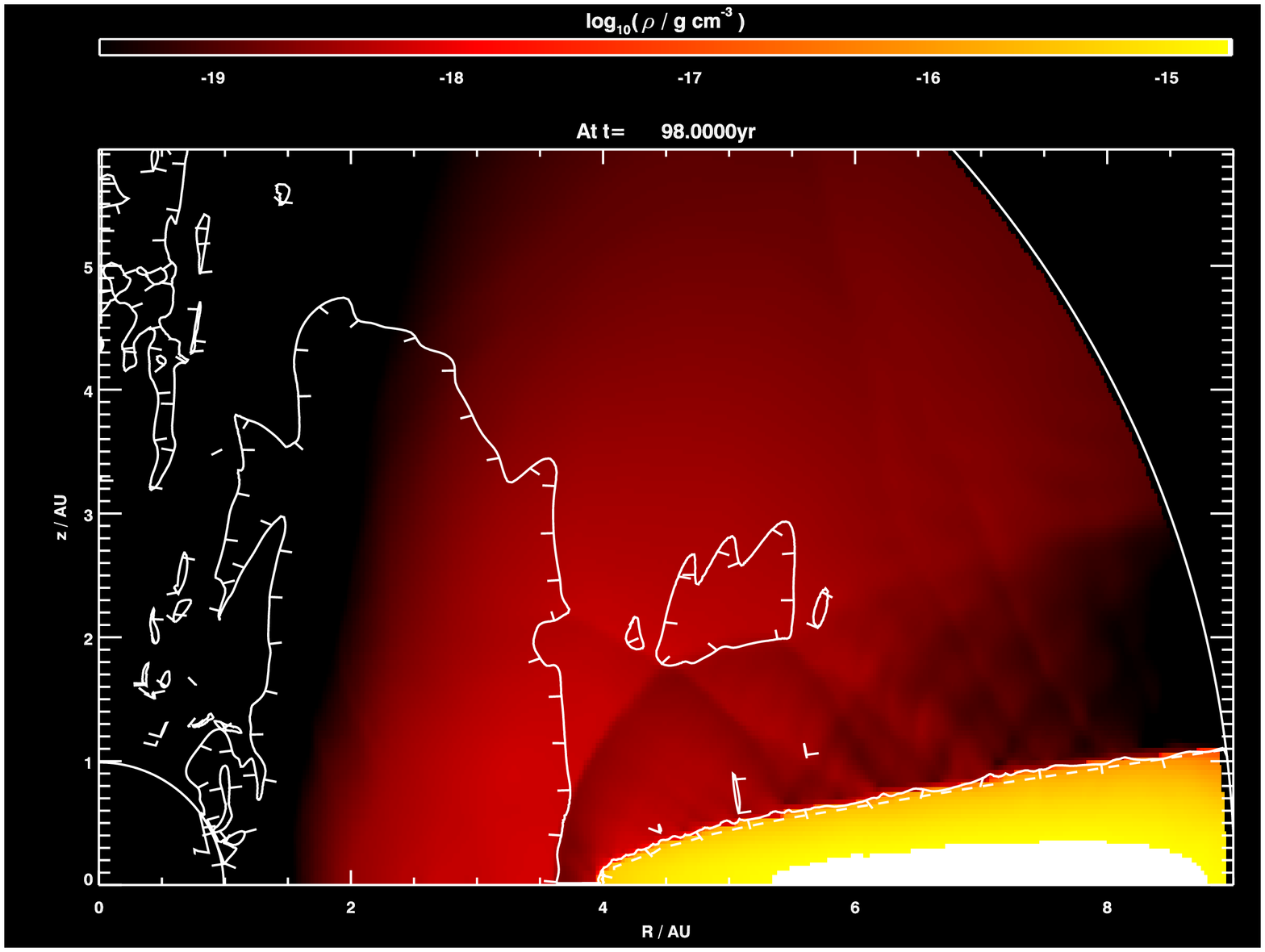}
        }

        \vspace*{5mm}

        \resizebox{\hsize}{!}{
        \includegraphics[angle=90]{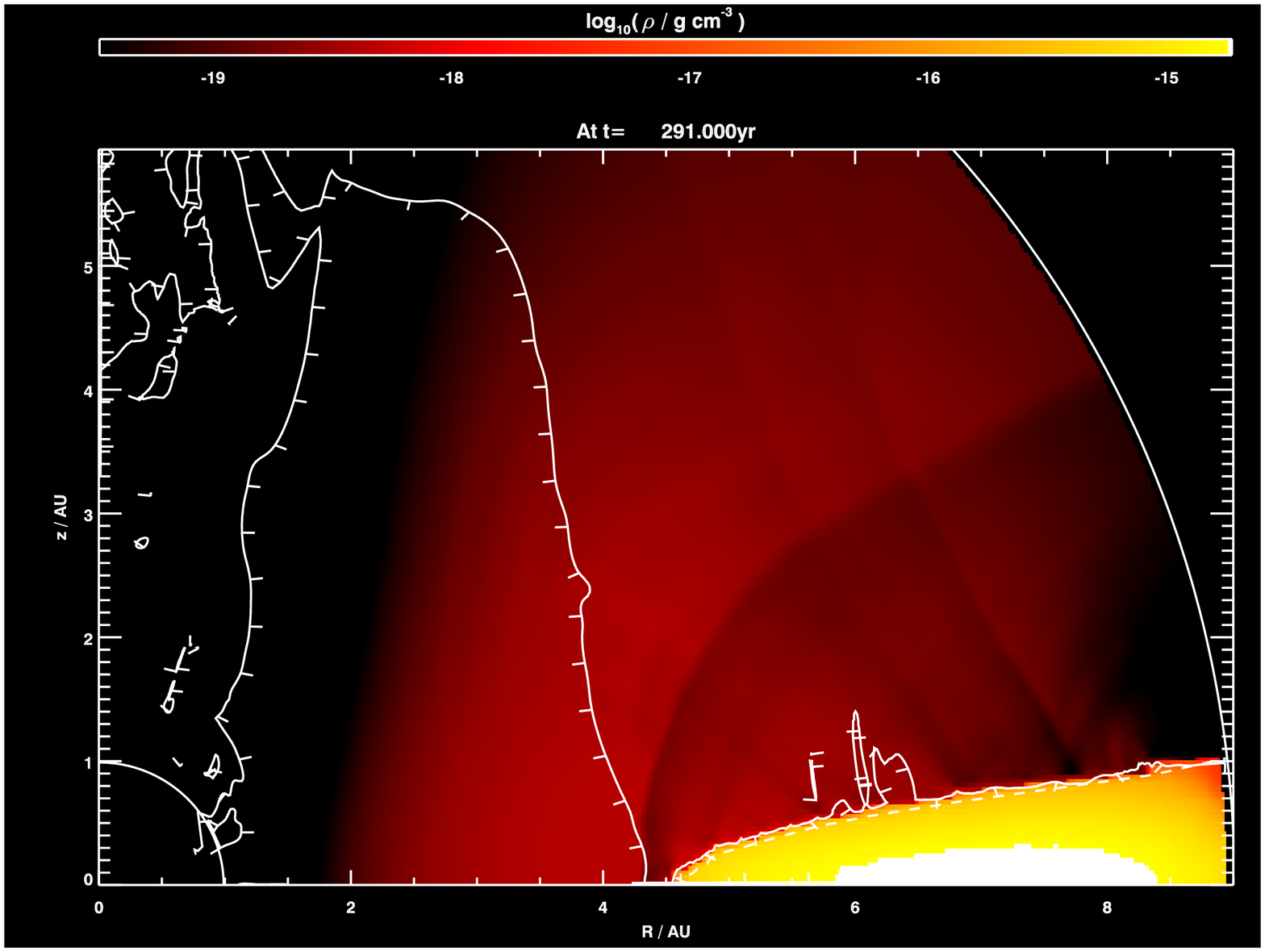}
        }
        \caption[Typical snapshot]{Snapshots of simulation {\sc Edge1} at $t=98$yr \& 291yr.  The plot is as in Fig.\ref{fig:snapshot}, but with the sonic surface plotted instead of velocity vectors.  The sonic surface is shown as a solid line, with tick marks indicating the sub-sonic side of the surface.  Note the ``messy'' velocity structure near to the ionization front in the second snapshot.  Comparison of the two snapshots also highlights the influence of the outer boundary condition on the outer region of the unperturbed disc.  By $t=291$yr ($\simeq 10.8$ orbital periods at the outer boundary) neutral gas has begun to "fall off" the disc at large radii due to the artificial pressure gradient introduced by the boundary condition, as can be seen by comparing the two snapshots.}
        \label{fig:sonic_surfs}
\end{figure}
In general the sonic surface is very close to, but slightly above, the ionization front (see Fig.\ref{fig:sonic_surfs}), suggesting that the flow is launched sub-sonically before rapidly becoming supersonic, as we expect for any steady wind.  However we note at this point that the velocity field near to the ionization front is not always stable.  This is because the ionization front is not coincident with the grid columns.  Consequently, as the front evolves across cell boundaries it can produces ``kinks'' in the front, which momentarily result in converging flows near to the front and are responsible for the striations visible in the flow snapshots.  As a result the velocity structure near to the front can be rather messy, while remaining smooth far from the front.  As long as the ionization front remains unresolved (as assumed in the algorithm) this problem will persist.  In practical terms this means that this effect is be present in all our simulations, regardless of the resolution.

A further feature of the algorithm is the treatment of material which flows in the $\theta$-direction, across different ray paths.  In our algorithm, the change in the number of ionized atoms is evaluated in such a way that any material flowing into a ray path is assumed to be neutral (see Equations \ref{eq:ioniz_rate}--\ref{eq:mass_enc}).  Thus flow across the front from the neutral to the ionized region is treated correctly.  However at higher latitudes, in the ionized region, this is incorrect and results in the ionization front being placed at too small a radius in some cases, resulting in some rather odd-looking front geometries at early times.  However in practice this is a transient effect which dies out as the flow becomes more stable, and only affects the simulations at early times.  Consequently we do not deem it to be a significant problem, as a similar time is needed for the other initial transients to disappear. 

\subsection{Convergence, boundary effects and other numerical issues}\label{sec:numerics}
The first computational consideration is the possible influence of numerical effects on the results obtained from the simulations.  Two specific issues are addressed: numerical convergence, and the influence of the artificial boundaries imposed upon the simulations.  In order to address the issue of numerical convergence we performed the {\sc ConvTest} simulation.  This simulation is identical to the fiducial model ({\sc Edge1}), but has double the spatial resolution in both dimensions.  In terms of the disc scale-height, $H/r\Delta\theta = 6.3$ in simulation {\sc Edge1}, and $H/r\Delta\theta = 12.7$ in {\sc ConvTest}.  To test convergence we evaluated the mass flux through a series of surfaces at fixed $r$ in the ionized region, multiplying by a factor of two to account for flow from both sides of the disc.  The surfaces used were not near to either the inner disc edge or the computational boundary in order to minimize their effects: surfaces at $r=5$, 6, 7, \& 8AU were used.  The quantity $\dot{M}(<r)$ should be independent of resolution, and so it is instructive to study the ratio of the different mass-fluxes between the two simulations.  The ratio of mass-fluxes ({\sc Edge1}/{\sc ConvTest}), time-averaged over $t=0$ to $t=80$yr (approximately three outer orbital periods), was found to be $1.04\pm0.06$, $1.02\pm0.05$, $1.04\pm 0.05$ \& $1.06\pm0.06$ at radii of 5, 6, 7 \& 8AU respectively.  These results suggest that the solution is converged, as the discrepancies between the two simulations are smaller than the fluctuations within each individual simulation.  The results also indicate that the numerical scheme is accurate to around $\pm5$\%.  Runs at lower resolution suggest that the solutions remain converged to better than 20\% accuracy as long as the disc scale-height $H(R)$ is revolved into at least 3 grid cells, and that this accuracy improves as the radial range increases.  (This is due to $\Delta r$ being fixed: a radial length of $H$ is resolved into more grid cells at larger radius)  Thus for $H/R=0.05$ the minimum resolution requirement for a marginally converged solution is $\Delta\theta = \pi/200$ (i.e.~100 cells in the angular direction).  We adopted double this resolution wherever possible, but were forced to use this minimum resolution when considering a large radial range in order keep the CPU requirements to a reasonable level.  Consequently we are satisfied that the simulations are numerically converged, although we note that the numerical accuracy of the {\sc Profile} simulations is likely only $\pm 15$\%.

The next issue to address is the influence of the computational boundaries on the simulations.  Both of the angular boundaries are axes of symmetry, and so the use of reflective boundaries here is physically reasonable.  Further, {\sc zeus2d} is exact in its treatment of reflective boundary conditions, so they are treated correctly by the code.  We note that the angular momentum barrier is sufficiently large that, initial transients aside, no mass actually reaches the $R=0$ angular boundary.  Both of the radial boundaries are set to be outflow boundaries.  {\sc zeus2d} is exact in its treatment of outflow boundaries as long as the flow is supersonic and along grid columns.  However sub-sonic outflow is known to produce spurious reflection at the boundaries, and supersonic outflows which are not along grid columns can also be problematic.  We note than in our simulations accretion through the inner boundary is negligible (and is exactly zero aside from initial transients), and so the inner boundary condition has no effect on the results.

In order to study the influence of the outer radial boundary condition we performed the simulation {\sc Boundary}.  This simulation has exactly the same parameters and resolution as the fiducial model ({\sc Edge1}), but covers double the radial range.  Consequently it is possible to study the influence of the outer boundary in the fiducial model by comparing the region near to the boundary in simulation {\sc Edge1} to the same region in simulation {\sc Boundary} (where it is in the middle of the computational domain).  We evaluated the mass outflow rates of both ionized and cold material through surfaces at $r=7$, 8, \& 9AU in both simulations.  The effect of the boundary on the ionized material is to decelerate the flow somewhat near to the boundary: the flow rates at 7AU and 8AU are essentially equal in the two simulations (time-averaged ratios of $1.02\pm0.06$ and $1.01\pm0.06$), but the outflow rate is somewhat smaller at 9AU in {\sc Edge1} than in {\sc Boundary} ($0.89\pm0.06$).  We hypothesise that this is caused by spurious reflections at the boundary, due to the flow direction not being radial (as supersonic outflow along grid columns should be treated exactly).  However we overcome this issue simply by neglecting the region near to the boundary: in the following sections we do not evaluate any variables at cells which are in the outer 10\% of the grid.  The effect of the boundary on the neutral material is to introduce a radial pressure gradient across the outer boundary, leading to some outflow of material.  The region influenced by this grows with time (as seen in Fig.\ref{fig:sonic_surfs}), and is also larger in simulations with greater $H/R$.  This effect is not significant in most of the simulations, but care must be taken when analyzing simulations with large $H/R$ or which have run to very large problem time.

\subsection{Inner edge density}\label{sec:inner_edge}
Three simulations were run in order to study the evolution of the inner edge of the disc.  Our simple analytic argument (Equation \ref{eq:n_in}) suggests that the density at the ionization front at the disc midplane should scale as 
\begin{equation}\label{eq:nin_form}
n_{\mathrm {in}} = C \left(\frac{\Phi}{4\pi \alpha (H/R)_{\mathrm{in}} R_{\mathrm {in}}^3}\right)^{1/2} \, ,
\end{equation}
where $C$ is an order-of-unity constant.  The scaling constant accounts for the fact that the radial scale-height is not exactly equal to the vertical scale-height, and also for the attenuation of the radiation field at $R < R_{\mathrm {in}}$.  We now seek to compare this prediction to the results of the simulations.

A problem arises when considering the density at the ionization front in the simulations.  The ionization front is not resolved, and so the density must be measured in the grid cell adjacent to the boundary cell.  This value decreases as the front advances across each cell, and so for the same value of $R_{\mathrm {in}}$ (i.e.~location of the boundary cell) a spread of densities is observed.  However by averaging over a large number of timesteps the noise introduced is minimized.  Additionally, as mentioned in Section \ref{sec:res} above, the evaluation of ``single-cell'' quantities is always resolution dependent.  Therefore, while it is possible to fit the power-law index accurately we find that the value of the normalisation constant $C$ depends somewhat on the grid resolution.

\begin{table}
 \centering
  \begin{tabular}{|cccc|}
  \hline
Simulation & Time Interval & \multicolumn{2}{c|}{Best-fitting parameters} \\
 & yr & Power-law index $b$ & Normalisation $C$ \\
  \hline
{\sc Edge1} & 15--1000 & $-1.54$ & $0.56$ \\
{\sc Edge2} & 15--500 & $-1.48$ & $0.79$ \\
{\sc Edge3} & 10--500 & $-1.46$ & $0.64$ \\
 & & & \\
Mean & - & $-1.49$ & $0.66$ \\
\hline
\end{tabular}
\vspace*{16pt}
\caption[Inner edge density]{Best-fitting parameters to inner edge density.  The parameters are fit using a simple least-squares algorithm.  The final row shows the mean values from the three simulations.  The typical (1$\sigma$) uncertainties in the fits are $\pm0.1$ in the power-law index and $\pm25$\% in the normalisation constant.}\label{tab:n_in}
\end{table}
\begin{figure}
\centering
        \resizebox{\hsize}{!}{
        \includegraphics[angle=270]{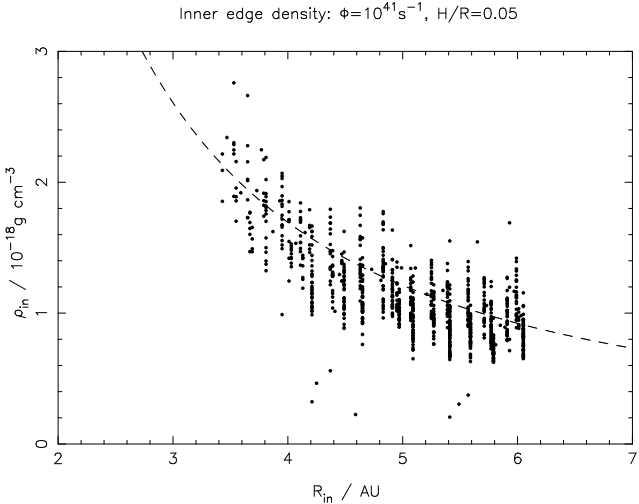}
        }
        \caption[Inner edge density]{Inner edge density plotted as a function of edge radius for simulation {\sc Edge1}.  Points are plotted every 1yr of problem time, from 15--1000yr.  A power-law fit is shown as a dashed-line, with index $b=-1.5$ and normalisation constant $C=0.6$.}
        \label{fig:n_in}
\end{figure}

In order to test Equation \ref{eq:nin_form} above we performed a fit to the data, using the form
\begin{equation}
\rho_{\mathrm {in}} = C m_{\mathrm H} \left(\frac{\Phi}{4\pi \alpha (H/R)}\right)^{1/2} R_{\mathrm {in}}^{b}\, .
\end{equation}
A two-parameter least-squares fit was performed to find the best-fitting power-law index $b$ and normalisation constant $C$.  The results of these fits are shown in Table \ref{tab:n_in} and Fig.\ref{fig:n_in}.  For simulations {\sc Edge2} and {\sc Edge3} a smaller range in problem time was used in order to lessen the influence of the outer boundary.  (The outer boundary is more significant in both of these simulations that in {\sc Edge1}, due to the larger $H/R$ in {\sc Edge2} and the faster progression of the inner edge in {\sc Edge3}.)  Due to the relatively small change in $R_{\mathrm {in}}$ over the course of the simulations there is a degeneracy between the two parameters, leading to rather large error bounds.  The best-fitting values, averaged over all three simulations, are $b=-1.49$ and $C=0.66$.  The typical (1$\sigma$) uncertainties in the fits are $\pm 0.1$ in $b$ and $\pm 25$\% in $C$.  Thus the results are entirely consistent with a power-law index of $-1.5$, as predicted analytically.  The {\sc ConvTest} simulation does not cover a sufficient radial range in $R_{\mathrm {in}}$ to constrain the power-law index well, giving a best-fitting value of $b=1.7\pm0.4$.  However when the index is set to $b=-1.5$ the best-fitting value of the normalisation constant is $0.83$.  Thus it seems that the value of $C$ is indeed somewhat resolution-dependent, but seems to lie in the range 0.5--1.0.  Note also that the determination of $C$ to be of order unity verifies our earlier assumption (Section \ref{sec:analytic_models}) that the thickness of the ionized layer is comparable to $H$, at least along the disc midplane.

\begin{figure*}
\centering
        \resizebox{\hsize}{!}{
        \includegraphics[angle=270]{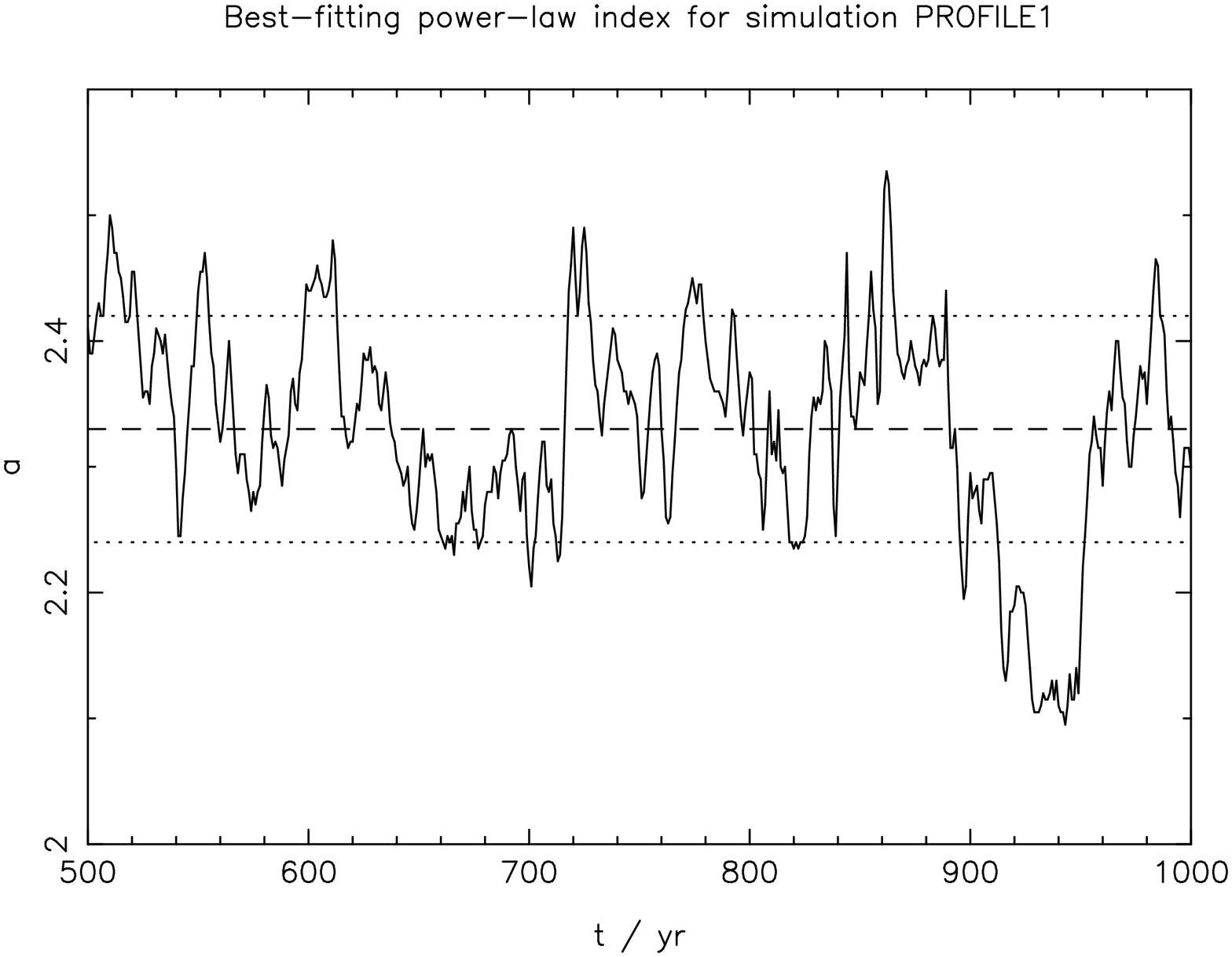}

        \hspace*{5mm}

        \includegraphics[angle=270]{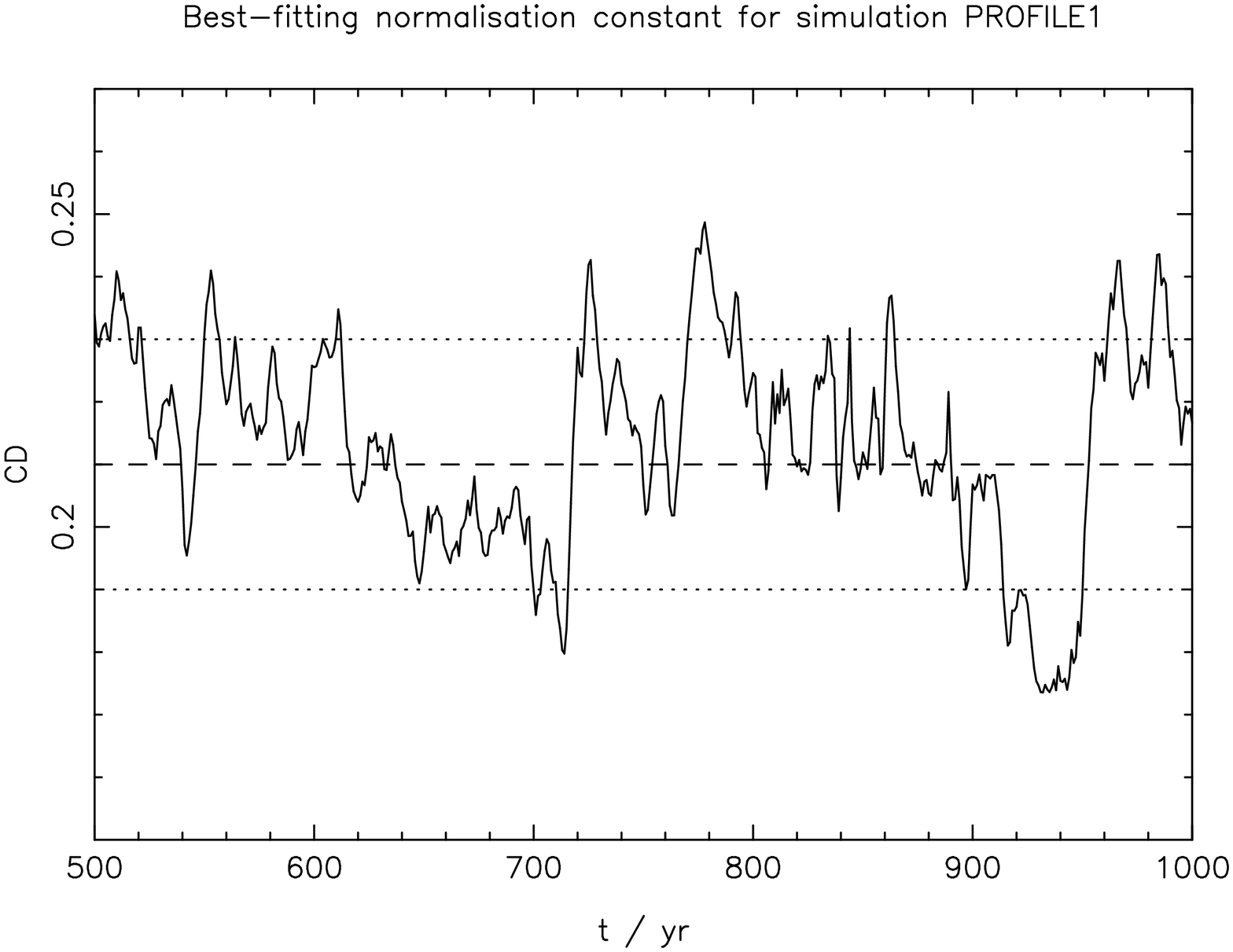}
        }
        \caption[Wind profile parameters]{Best-fitting parameters to the wind profile as a function of time for simulation {\sc Profile1}.  The left-hand panel shows the power-law index $a$, the right-hand panel the normalisation constant $(CD)$.  In both plots the mean value is shown as a dashed line, with the $\pm1\sigma$ error bounds shown as dotted lines.}
        \label{fig:mdot_params}
\end{figure*}

\subsection{Launch velocity}
A number of simulations were run to try to constrain the launch velocity and determine the value of the constant $D$ in Equation \ref{eq:launch_vel}.  However, as mentioned above, the velocity field near to the ionization front fluctuates and it was not possible to obtain good constraints on the launch speed.  In general it seems that the launch speed is approximately constant, as there is no significant variation with radius.  However the fluctuations in the velocity at any given radius are large.  In general the launch speed seems to be slightly sub-sonic, although it can become supersonic at times.  The best-fitting values of $D$ show a large scatter, and cannot be constrained better than to say that the launch speed is, as expected, of order the sound speed.  It is necessary to consider ``many-cell'' variables in order to obtain a more accurate result.

\subsection{Wind profile}
\begin{table*}
 \centering
  \begin{tabular}{|cccccc|}
  \hline
Simulation & Time Interval & $H/R$ & $\Phi$ & \multicolumn{2}{c|}{Time-averaged parameters} \\
 & yr & & $10^{41}$s$^{-1}$ & $a$ & $(CD)$ \\
  \hline
{\sc Profile1} & 500--1000 & 0.05 & 1.0 & $2.33\pm0.09$ & $0.21\pm0.02$ \\
{\sc Profile2} & 250--500 & 0.1 & 1.0 & $4.50\pm0.13$ & $0.60\pm0.02$ \\
{\sc Profile3} & 250--500 & 0.05 & 10.0 & $2.49\pm0.08$ & $0.26\pm0.02$ \\
{\sc Profile4} & 250--500 & 0.075 & 1.0 & $3.38\pm0.18$ & $0.43\pm0.03$ \\
\hline
\end{tabular}
\vspace*{16pt}
\caption[Mass-loss profile parameters]{Time averages of the scaling parameters for the mass-loss profile, evaluated from each simulation.  $a$ is the power-law index, and $(CD)$ is the normalisation constant.  The errors quoted are (1$\sigma$) standard deviations.}\label{tab:mdot_params}
\end{table*}

\begin{figure*}
\centering
        \resizebox{\hsize}{!}{
        \includegraphics[angle=270]{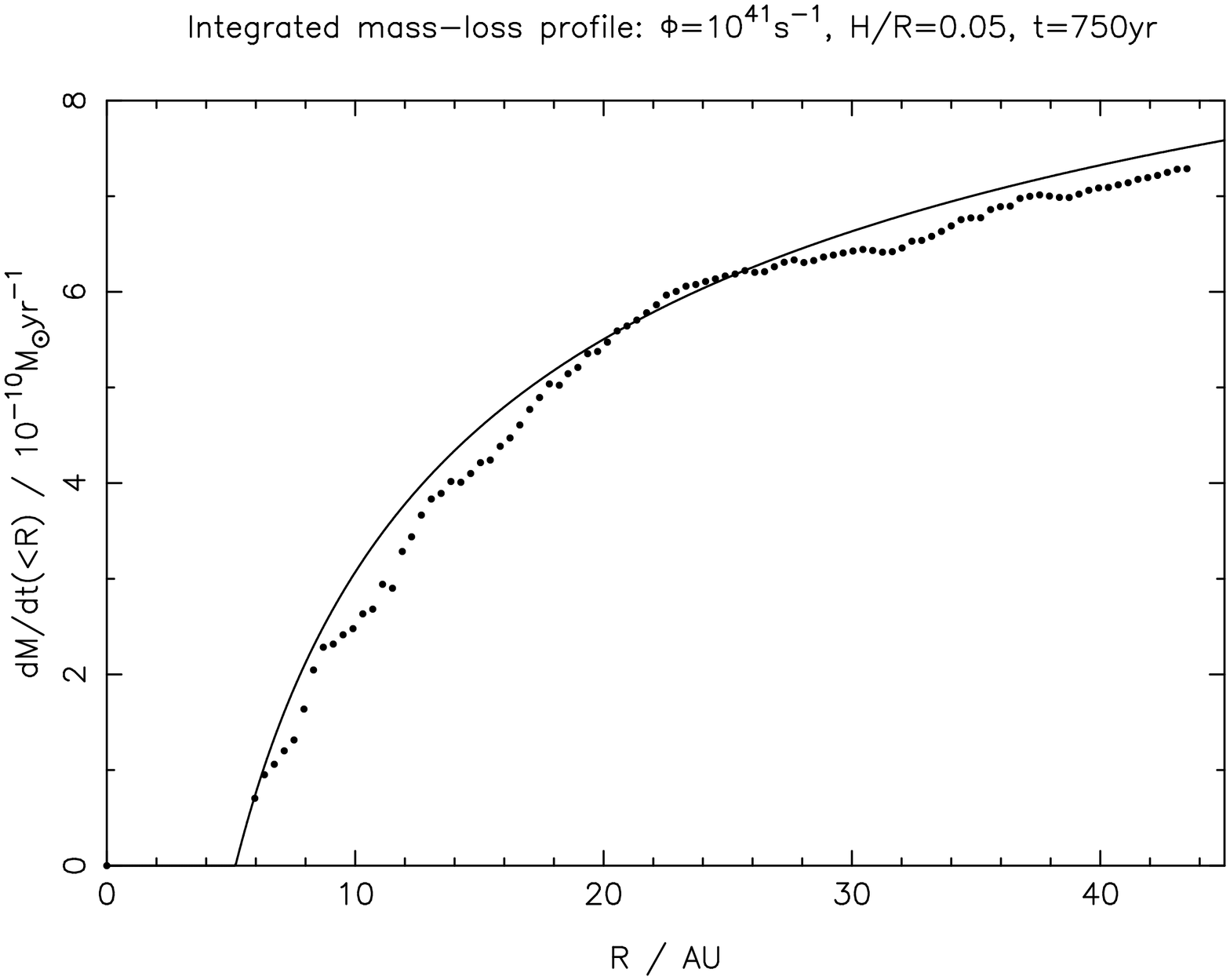}

        \hspace*{5mm}

        \includegraphics[angle=270]{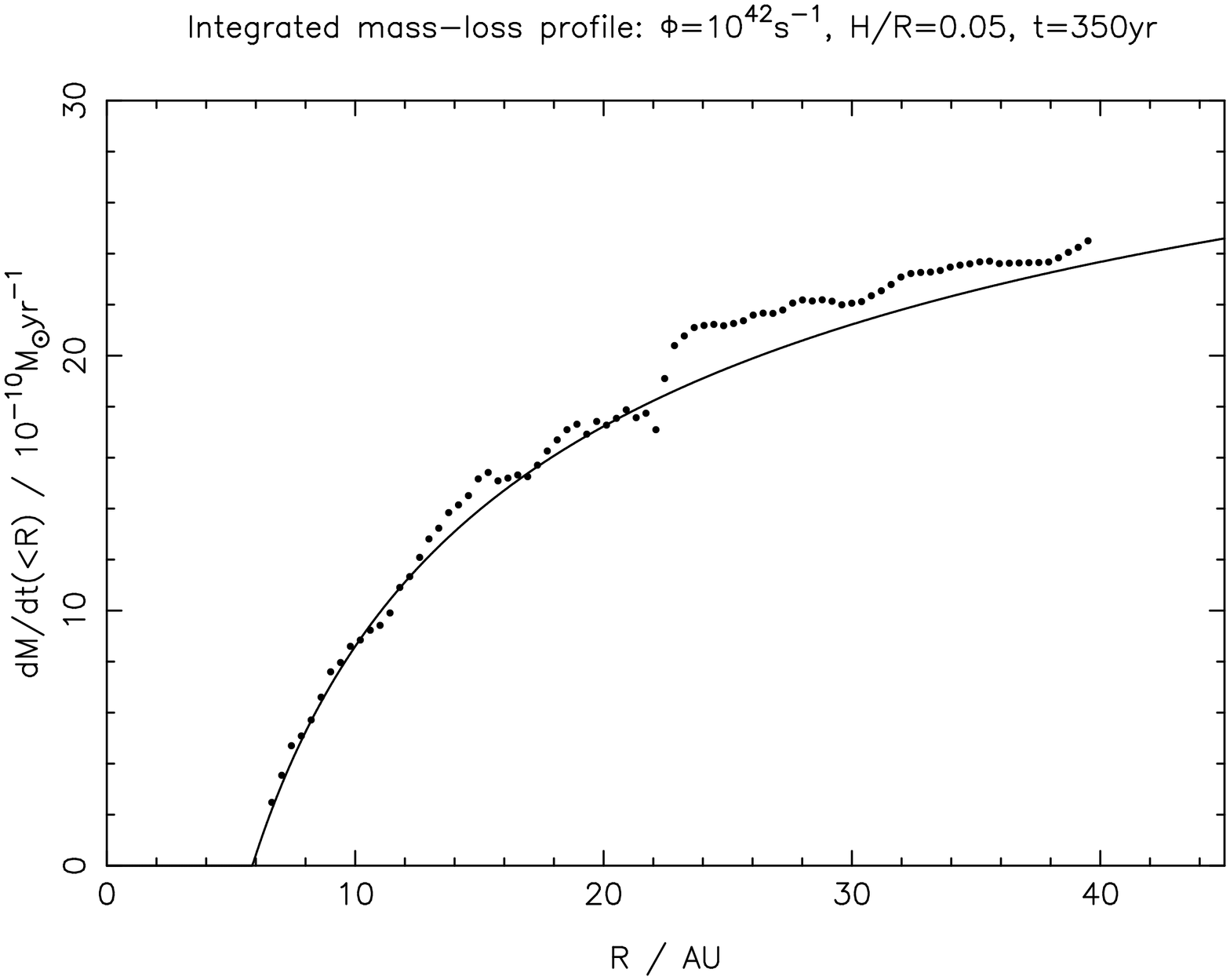}
        }
        \caption[Fits to mass-loss profiles]{Snapshots of the mass-loss profile from the simulations.  The left-hand panel shows the results for simulation {\sc Profile1} ($\Phi=10^{41}$s$^{-1}$, $H/R=0.05$) at $t=750$yr, the right-hand panel simulation {\sc Profile3} ($\Phi=10^{42}$s$^{-1}$, $H/R=0.05$) at $t=350$yr.  In both plots the dots are the data from the simulations, with the solid line showing the analytic form of the mass-loss profile (Equation \ref{eq:mout_anal}) for $a=2.42$ and $(CD)=0.235$: in both cases the fit is extremely good. }
        \label{fig:mdot_fits}
\end{figure*}
In order to study the wind profiles produced by the model we conducted the {\sc Profile} series of simulations.  These simulations cover a large radial range, enabling the study of the wind profile over a wider range of radii than possible with the higher resolution {\sc edge} simulations.  The method adopted to analyse the results of these simulations was to study the mass outflow rate across given radial shells at each timestep.  At radii outside the inner disc edge the outflowing ionized mass through a shell of constant $r$ comes from the disc surface inside a cylindrical radius $R = r\sin \theta_{\mathrm f}$ (where $\theta_{\mathrm f}$ is the angular coordinate of the ionization front).  Consequently by sampling $\dot{M}(<r)$ at many radii it is possible to study the mass-loss profile $\dot{\Sigma}(R)$.

In order to study the simulations in more detail we compared our analytic prediction for $\dot{M}(<R)$ (derived in Section \ref{sec:analytic_models}) with the values obtained from the simulation.  In general we find that a power-law form for the shape function
\begin{equation}
f(x) = x^{-a}
\end{equation}
provides a good fit to the profile, although the values of $a$ varies for different values of $H/R$ (i.e.~we fit a different shape function for different $H/R$).  Substituting this form into the the analytic expression for $\dot{M}(<R)$ (Equation \ref{eq:mout_anal}) and integrating gives
\begin{eqnarray}
\dot{M}(<R_{\mathrm {out}}) = 1.73\times10^{-9} \, \frac{CD}{a-2} \, \mu \, \left(\frac{\Phi}{10^{41}\mathrm s^{-1}}\right)^{1/2}  \left(\frac{R_{\mathrm {in}}}{3\mathrm{AU}}\right)^{1/2}
\nonumber \\
\times \left(\frac{H/R}{0.05}\right)^{-1/2}  \left[1 - \left(\frac{R_{\mathrm {in}}}{R_{\mathrm {out}}}\right)^{a-2}\right] \, \mathrm M_{\odot}\mathrm{yr}^{-1} \, ,
\end{eqnarray}
for $a \ne 2$.  Therefore by comparing to the numerical results it is possible to obtain values for the combined scaling constant $(CD)$ and the power-law index $a$.  Our procedure for fitting these parameters is as follows.

At each timestep we evaluate $\dot{M}(<R)$ at a large number of different radii on the grid.  The mass outflow rate is evaluated as
\begin{equation}
\dot{M}(<r_i) = 2\sum_j \rho_{i,j} u_{r(i,j)} \left(2\pi \Delta r_i^2 \sin \theta_j \Delta \theta \right) 
\end{equation}
in the ionized region.  Here $u_{r(i,j)}$ is the radial component of the fluid velocity at cell $(i,j)$, the last term represents the area of the outer radial surface of each cell, and the initial factor of 2 accounts for flow from both sides of the disc.  The smallest value of $r_i$ used was 0.5AU larger than the inner edge radius at that timestep, and we do not evaluate $\dot{M}(<r_i)$ in the region where the outer boundary becomes significant (the outer 10\% of the grid for most models, but somewhat larger where $H/R>0.05$).  We then solve for the values of $a$ and $(CD)$ which give the best-fit to a straight line in the $\dot{M}(<R)$/$\left[1 - \left(R_{\mathrm {in}}/R\right)^{a-2}\right]$ plane.  These values fluctuate somewhat with time, but by studying a large number of timesteps it is possible to obtain best-fitting values.  In each case it is necessary to wait until the simulation stabilises before studying the mass-loss profile: typically this takes around one outer orbital time.  The evolution of the parameters with time for the {\sc Profile1} simulation are shown in Fig.\ref{fig:mdot_params}, and the mean values of the constants for each simulation are shown in Table \ref{tab:mdot_params}.

In general we find that a single power-law index and a single normalisation constant provide a good fit to the mass-loss profile for fixed $H/R$.  For $H/R=0.05$ the best-fitting values are $a=2.42\pm0.09$ and $(CD)=0.235\pm0.02$: examples of this fit are shown in Fig.\ref{fig:mdot_fits}.  The single value obtained confirms that the mass-loss rate scales as $\Phi^{1/2}$, which in turn confirms that the flow is ``recombination-limited'' \citep[see][]{holl94}.  Note also that this value of $(CD)$ is consistent with $C\simeq0.6$, as estimated from the inner edge density, if $D\simeq0.4$.  This suggests that the typical launch velocity is around 0.4 times the sound speed, a value consistent with numerical simulations of the photoevaporative wind driven by the diffuse radiation field \citep{font04}.

\begin{figure}
\centering
        \resizebox{\hsize}{!}{
        \includegraphics[angle=270]{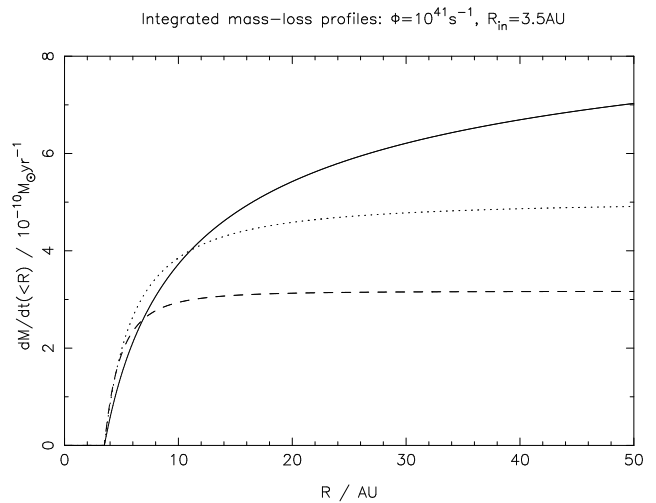}
        }
        \caption[Variation of mass-loss profile with $H/R$.]{Best-fitting mass-loss profiles for various values of $H/R$.  The profiles are evaluated for fixed values of $\Phi=10^{41}$s$^{-1}$ and $R_{\mathrm {in}}=3.5$AU, using the best-fitting parameters listed in Table \ref{tab:mdot_params}.  The solid line shows the profile for $H/R=0.05$, the dotted line $H/R=0.075$, and the dashed line $H/R=0.10$.  While the form of the profiles varies somewhat with $H/R$, the integrated mass-loss rates are not especially sensitive to the disc thickness.}
        \label{fig:HR_mdot}
\end{figure}

However the power-law index is rather sensitive to the value of $H/R$, as seen in Table \ref{tab:mdot_params}.  This is not entirely surprising, as the disc density decreases exponentially with $(z/H)^2$, meaning that the density at which ionization balance occurs is rather sensitive to the disc thickness.  This obviously has a knock-on effect for the integrated mass-loss rates, with thicker discs resulting is less efficient mass-loss at radii beyond the inner disc edge.  However while the shape of the mass-loss profile in terms of $\dot{\Sigma}(R)$ varies significantly with the value of $H/R$, the effect on the integrated mass-loss rate is not especially strong.  This is illustrated in Fig.\ref{fig:HR_mdot}, which shows the best-fitting mass-loss profiles for three different values of $H/R$.  Larger values of $H/R$ result in mass-loss profiles which are rather more concentrated towards the inner disc edge, but the total integrated mass-loss rates differ only by a factor of 2--3.

\section{Discussion}\label{sec:disc}
In order to construct these models we have made a number of simplifications and approximations, and now seek to address their significance.  The most important of these is the use of the on-the-spot (OTS) approximation to simplify the radiative transfer problem.  Throughout the simulations we have adopted Case B recombination coefficient, which has a value of $\alpha_{\mathrm B}=2.6\times10^{-13}$cm$^3$s$^{-1}$ \citep{allen}.  Adopting the Case A value ($\alpha_{\mathrm A}=4.2\times10^{-13}$cm$^3$s$^{-1}$) has a negligible effect on the results, as the density in the ionization balance equation depends on $\alpha^{-1/2}$.  However we must also consider the possibility that recombination photons can be absorbed non-locally: were this to be significant the OTS approximation would fail.  In the simulations the ionized region is always optically thin to ionizing photons.  (In fact, the density in the ionized region is sufficiently low that it would not be optically thick to ionizing photons even if the gas were entirely neutral.)  Similarly, the density in the cold disc is sufficiently large that the mean free path of an ionizing photon is always much smaller than the length of a grid cell.  Consequently, if we consider recombinations at a given point on the front we see that any photons emitted in an upward direction escape the system, whereas any photons emitted downwards are absorbed locally.  If we look at the geometry of the ionization front (see Figs.~\ref{fig:snapshot} \& \ref{fig:sonic_surfs}), we see that the front approaches the line-of-sight from the source asymptotically at large radii.  Therefore there is only a very small solid angle which permits recombination photons to be absorbed elsewhere along the front, and their influence is negligible by comparison to the direct field.  Consequently we are satisfied that the OTS approximation is valid for this geometry.

A further simplification related to the the use of the OTS approximation is that recombination photons produced in the flow region are similarly neglected.  This diffuse field is the field that drives the disc wind in the models of \citet{holl94}, and so by comparison to their work it is possible to estimate its contribution.  \citet{holl94} found that the diffuse field produces around 10\% of the flux of the direct field in the static region.  Here the diffuse field must come from ionized material that is flowing away from the disc, and so may be somewhat less efficient than in the static case.  However if we neglect the motion of the flow it seems that neglecting the diffuse field produced by recombinations in the flow results in our model under-estimating the radiation field at the front by around 10\%.  We suggest, however, that a more detailed radiative transfer calculation is needed in order to confirm this.  We note that both of the simplifications of the OTS approximation result in our model under-estimating the ionizing flux at the ionization front, and therefore suggest that our mass-loss rates can be considered as lower-limits to the true mass-loss rates due to direct photoevaporation.  Moreover the wind rate scales as $\Phi^{1/2}$, and so a small increase in the effective value of $\Phi$ does not have a strong influence on the total mass-loss rate.

The second simplification we have made is the equation of state adopted in the ``boundary'' cells (Equations \ref{eq:EoS} \& \ref{eq:c_bc}).  This approximation may be significant, as the treatment of the boundary cell can have a significant effect on the flow solution.  As mentioned above, we treat the boundary cells in such a way that any material flowing ``laterally'' is assumed to be neutral.  In practice the pressure gradients across the ionization front are upwards, from ``neutral'' to ``ionized'', and so this should predict the correct behaviour.  However with no model with which to compare our results it is difficult to quantify the accuracy of this approximation.  As seen in Section \ref{sec:Spitzer_test}, comparison to the Spitzer solution suggests an accuracy of around 5\%, but we also note this is not an especially good test of the code's applicability to the direct photoevaporation problem.  Comparison of our results with independently obtained solution in the future will be a valuable test of the accuracy of the code.

As noted in Section \ref{sec:analytic_models}, our model neglects any possible dust opacity between the star and the inner disc edge.  The presence of dust in the ``inner hole'' could attenuate the flux of ionizing photons reaching the disc, and consequently reduce the wind rate produced by direct photoevaporation.  However it is not clear whether or not the viscous accretion of the inner gas disc will also remove the dust, and a detailed treatment of the gas-grain dynamics during the transition phase is beyond the scope of this work.  Some insight can be gained from the work of \citet{tcl05}, which considers the differential motions of dust and gas in a simple photoevaporating disc model.  In general they find that dust is rapidly accreted with the gas when the inner disc drains, but there is a complicated dependence on the various model parameters.  The  results of \citet{tcl05} therefore suggest that the effect of dust opacity in the inner part of the disc will be small, but we note that this issue warrants further study.

As seen above, the mass-loss profiles obtained by the model are rather sensitive to the disc scale-height.  We note also that real TT discs are expected to show significant ``flaring'' \citep[e.g.][]{kh87}, and are not generally consistent with a constant $H/R$ ratio.  A flaring disc would show a slower decrease in the incidence term $\sin \beta$ (Equation \ref{eq:base_den}) with $R$ than a disc with constant $H/R$.  Consequently a flaring disc is expected to show greater mass-loss at large radii than seen in our models.  However once again this suggests that our models provide a robust lower-limit to the mass-loss rates produced by direct photoevaporation.

Lastly, it is instructive to compare the mass-loss profile we obtain to that predicted by diffuse photoevaporation.  In the diffuse model the wind profile takes the form
\begin{equation}
\dot{\Sigma}_{\mathrm {wind}} \propto R^{-5/2} \, , 
\end{equation}
with the profile normalised at $R_{\mathrm g}$ regardless of the instantaneous location of the inner disc edge \citep{holl94}.  Therefore as the inner disc edge moves outward to $R_{\mathrm {in}} > R_{\mathrm g}$, the integrated mass-loss rate decreases significantly.  However when direct photoevaporation is considered we find that while the decline in the radial power-law is similar (for $H/R=0.05$), the profile is instead normalised at the instantaneous inner disc edge (see Equation \ref{eq:prof_form}).  As discussed in Section \ref{sec:analytic_models}, the form of the direct wind is qualitativelty similar to that found by \citet{holl94} for the case of a strong stellar wind.  This form results in a significantly larger $\dot{M}$ as the inner disc edge evolves outwards, as the integrated mass-loss rate scales with $R_{\mathrm {in}}^{1/2}$ (Equation \ref{eq:mout_anal}).  Additionally, the more effective radiative transfer in the case of direct photoevaporation results in a mass-loss rate that is larger than that expected in the diffuse case, by a factor of $\simeq 5$--10 at $R_{\mathrm {in}} \sim R_{\mathrm g}$.  Therefore we expect that direct photoevaporation is much more efficient than diffuse photoevaporation in dispersing the outer part of the disc.  This provides a possible solution to the ``outer disc problem'' of the original UV-switch model \citep{cc01}.  We explore the effect of this modified mass-loss profile on the evolution of TT discs in a companion paper (Paper II).

\section{Summary}\label{sec:summary}
We have highlighted a significant flaw in disc evolution models which incorporate photoevaporation, namely that such models neglect the direct radiation field after the inner disc has been drained.  We have modelled the photoevaporative wind produced by the direct field, firstly using a simple analytic treatment and then using detailed numerical hydrodynamics.  Our models show that once the inner disc has drained the wind due to the direct field dominates over that due to the diffuse field.  The presence of such a wind leads to a significantly faster dispersal of the outer disc than seen in existing models of disc photoevaporation \citep[e.g.][]{cc01}, as the integrated mass-loss rate scales with $R_{\mathrm {in}}^{1/2}$ as $R_{\mathrm {in}}$ grows (cf.~$R_{\mathrm {in}}^{-1/2}$ in the diffuse case, \citealt{holl94}).  We suggest that this may be a solution to the ``outer disc problem'' in these models.  Our analysis results in a simple form for the mass-loss due to direct photoevaporation, and we explore the effects of this wind on models of disc evolution in a companion paper (Paper II).

\section*{Acknowledgements}
We thank David Hollenbach for a comment which initiated this investigation.  We thank Doug Johnstone, Andreea Font and Ian McCarthy for providing numerical results from their hydrodynamic models.  We also thank Doug Johnstone for a useful referee's report, which drew our attention to issues we had not previously considered.  RDA acknowledges the support of a PPARC PhD studentship.  CJC gratefully acknowledges support from the Leverhulme Trust in the form of a Philip Leverhulme Prize.  Parts of this work were supported by NASA under grant NNG05GI92G from the Beyond Einstein Foundation Science Program.


\label{lastpage}

\end{document}